\documentclass[12pt]{iopart}

\usepackage{graphicx,psfrag}
\begin{document}

\title[Massive Charged Scalar QNMs of RN Black Hole Surrounded by Quintessence]{Massive Charged Scalar Quasinormal Modes of Reissner-N\"ordstrom Black Hole Surrounded by Quintessence}

\author{Nijo Varghese and V C Kuriakose}

\address{Department of Physics, Cochin University of Science and
Technology,

Cochin - 682 022, Kerala, India} \ead{nijovarghese@cusat.ac.in and
vck@cusat.ac.in}
\begin{abstract}
We evaluate the complex frequencies of the normal modes for the
massive charged scalar field perturbations around a
Reissner-N\"ordstrom black hole surrounded by a static and
spherically symmetric quintessence using third order WKB
approximation approach. Due to the presence of quintessence,
quasinormal frequencies damp more slowly. We studied the variation
of quasinormal frequencies with charge of the black bole, mass and
charge of perturbating scalar field and the quintessential state
parameter.

\end{abstract}

\maketitle

\section{Introduction}
It is now well understood that different kinds of perturbations in
the geometry of a black hole can excite certain combination of its
characteristic complex frequencies of the normal mode, called
quasinormal modes(QNMs), whose real part represents the ring down
frequency and imaginary part, the decay time and they are
independent of the initial perturbation, but depends only on the
parameters of black hole. The concept of QNM is put forward by
Vishweshwara\cite{ref-4} and since then the quasinormal spectrum of
black holes has been extensively studied for a great verity of black
hole back grounds and perturbation fields because of its
astrophysical and other theoretical interests. Astrophysical
interests are associated with their relevance in gravitational wave
analysis\cite{ref-3}. The QNMs of black hole are expected to be
detected by the future gravitational wave detectors such as
LISA\cite{ref-2}, and give an opportunity to explore the properties
of black holes directly. Apart from the observational interest of
detection of quasinormal ringing by gravitational wave detectors,
study of QNM found significance in AdS/CFT
correspondence\cite{ref0}, black hole area quantization and Loop
Quantum Gravity\cite{ref-1}.

The idea that our universe is in a phase of accelerated expansion
rather than holding steady, is strongly supported by a set of recent
interlocking observations such as studies of distant
supernovae\cite{refa}, type 1a supernova\cite{refb} and cosmic
microwave background radiation anisotropy\cite{refc}, indicating the
presence of some mysterious form of repulsive gravity called dark
energy. In order to explain the nature of dark energy several models
were proposed(for a recent review see\cite{refl}). The simplest
option for this dark energy is Einstein's cosmological
constant\cite{refd} with a constant equation of states(EOSs)
$\epsilon=-1$ but it needs extreme fine tuning to account for the
observations. The second is the dynamical scalar field models like
quintessence\cite{refe}, k-essence\cite{reff} and
phantom\cite{refg}, in which the EOSs of dark energy changes with
time. Among them, the quintessence is the most natural model, and
can give the EOSs with $-1\leq\epsilon\leq-1/3$.

Because of relevance of the two topics, some studies of QNMs of
black hole were started in the presence of quintessence, after
Kiselev\cite{ref1} derived the exact solution of Einstein equation
for quintessential matter surrounding a black hole. QNMs of
Schwarzschild black hole is studied for scalar\cite{refh},
gravitational\cite{refi}, electromagnetic\cite{refj} and massless
Dirac field\cite{refk} perturbations in the presence of
quintessence. Our work extend the studies to charged black hole
space times. Charged scalar QNMs of Reissner-Nordstrom black hole
were studied in\cite{ref6,ref8}. In this paper we considered the
perturbation of Reissner-Nordstrom black hole encircled by
quintessence for massive charged scalar field perturbations and
study the dependence of quasinormal spectrum upon different
parameters of the problem such as charge of the black bole, mass and
charge of perturbating scalar field and the quintessential state
parameter.

\section{Massive charged scalar field around a Reissner-Nordstrom black hole surrounded by quintessence} Kiselev derived a
static spherically symmetric exact solution of Einstein equations
for quintessential matter surrounding a charged black hole under the
condition of additivity and linearity in energy momentum
tensor\cite{ref1}. The metric can be expressed in the form

\begin{equation}
\ ds^{2}=-f(r)dt^{2}+f(r)^{-1}dr^{2}+r^{2}d\Omega ^{2}, \
\label{eqn1}
\end{equation}

where

\[
f(r)=(1-\frac{2M}{r}+\frac{Q^{2}}{r^{2}}-\frac{c}{r^{3\epsilon
+1}})\\ \ \ \ \ and \ \ \ \ \ \ d\Omega ^{2}=(d\theta ^{2}+\sin
^{2}\theta d\phi ^{2}),
\]

with $M$ and $Q$ mass and charge of the black hole. $\epsilon $ is
the quintessential state parameter and $c$ is the normalization
factor which depends on $\rho _{q}=\frac{-c}{2}\frac{3\epsilon
}{r^{3(1-\epsilon )}}$,the density of quintessence. The Klein-Gordon
equation describing the evolution of massive charged scalar
perturbation field outside a charged black hole is given
by\cite{ref2}
\begin{equation}
\Phi _{;\mu \nu }g^{\mu \nu }-ieA_{\mu }g^{\mu \nu }(2\Phi _{;\nu
}-ieA_{\nu }\Phi )-ieA_{\mu ;\nu }g^{\mu \nu }\Phi =m^{2}\Phi ,
\label{eqn2}
\end{equation}

where $A_{\mu }$ is the electromagnetic potential and $e$ is the
charge of the scalar field. Representing the charged scalar field in
to spherical harmonics

\begin{equation}
\Phi =\frac{1}{r}\sum\limits_{l,m}\eta _{m}^{l}(t,r)Y_{l}^{m}(\theta
,\phi ), \label{eqn3}
\end{equation}

the wave equation becomes \qquad

\begin{equation}
\eta _{,tt}-2ieA_{t}\eta _{,t}-\eta _{,r^{\ast }r^{\ast }}+f(r)\left[ \frac{%
l(l+1)}{r^{2}}+\frac{f(r)^{^{\prime }}}{r}+m^{2}\right] \eta
-e^{2}A_{t}^{2}\eta =0, \label{eqn4}
\end{equation}

where we used the coordinate transformation defined by

\begin{equation}
dr^{\ast }=\frac{dr}{f(r)}. \label{eqn5}
\end{equation}

The electromagnetic potential $A_{t}=C-\frac{Q}{r}$ ,where $C$ is a
constant. We define

\begin{equation}
\eta =e^{-ieCt}\Psi, \label{eqn6}
\end{equation}

to avoid the physically unimportant quantity $C$. where $\Psi$ is
the radial part of perturbation variable taken to have time
dependence $\e^{-i\omega t}$. Now the radial perturbation equation
can be written as

\begin{equation}
\Psi _{,r^{\ast }r^{\ast }}+ \Theta \Psi =0 \label{eqn7}
\end{equation}

 where $\Theta=\omega ^{2}-V^{2}$ and $V$ is the scattering potential, which is a function of frequency of perturbation, $\omega$ and is given by

\begin{equation}
V=f(r)\left[ \frac{l(l+1)}{r^{2}}+\frac{2M}{r^{3}}-\frac{2Q^{2}}{r^{4}}+%
\frac{c(3\epsilon +1)}{r^{3\epsilon +3}}+m^{2}\right] +\frac{2eQ\omega }{r}%
-e^{2}\frac{Q^{2}}{r^{2}} \label{eqn8}
\end{equation}

\section{Evaluation of Quasinormal modes}
We use the third order WKB approximation method to determine the
complex normal mode frequencies of black hole, a semi analytic
method originally developed by Schutz and Will\cite{ref3} for the
lowest order. Later Iyer and Will\cite{ref4} carried this approach
to third WKB order and Konoplya\cite{ref5} to sixth order to get
more accurate results.

It gives a simple condition which will lead to discreet, complex
values for the normal mode frequencies.
\begin{equation}
\frac{i\Theta_{0}}{\sqrt{2\Theta_{0}^{^{\prime \prime }}}}-\Lambda (n)-\Omega (n)=n+%
\frac{1}{2}, \label{eqn9}
\end{equation}

where $\Lambda$ and $\Omega$ are higher order terms given by
\begin{equation}
\Lambda(n) =\frac{i}{(2\Theta_{0}^{^{\prime \prime
}})^{1/2}}[\frac{1}{8}\left(
\frac{\Theta_{0}^{(4)}}{\Theta_{0}^{^{\prime \prime }}}\right) \left( \frac{1}{4}%
+\alpha ^{2}\right) -\frac{1}{288}\left( \frac{\Theta_{0}^{^{\prime
\prime \prime }}}{\Theta_{0}^{^{\prime \prime }}}\right) ^{2}\left(
7+60\alpha ^{2}\right) ], \label{eqn10}
\end{equation}
\qquad \qquad \qquad \qquad \bigskip
\begin{eqnarray}
\Omega  &=&\frac{\alpha }{(2\Theta_{0}^{^{\prime \prime }})}\{\frac{5}{6912}%
\left( \frac{\Theta_{0}^{^{\prime \prime \prime }}}{\Theta_{0}^{^{\prime \prime }}}%
\right) ^{4}\left( 77+188\alpha ^{2}\right) -\frac{1}{384}\left( \frac{%
\Theta_{0}^{^{\prime \prime \prime }2}\Theta_{0}^{^{(4)}}}{\Theta_{0}^{^{\prime \prime }3}}%
\right)   \nonumber \\
&&\left( 51+100\alpha ^{2}\right) +\frac{1}{2304}\left( \frac{\Theta_{0}^{(4)}}{%
\Theta_{0}^{^{\prime \prime }}}\right) ^{2}\left( 67+68\alpha ^{2}\right) +\frac{1%
}{288}\left( \frac{\Theta_{0}^{^{\prime \prime \prime }}\Theta_{0}^{^{(5)}}}{%
\Theta_{0}^{^{\prime \prime }2}}\right)   \nonumber \\
&&\left( 19+28\alpha ^{2}\right) +\frac{1}{288}\left( \frac{\Theta_{0}^{^{(6)}}}{%
\Theta_{0}^{^{\prime \prime }}}\right) \left( 5+4\alpha ^{2}\right)
\}. \label{eqn11}
\end{eqnarray}

where
\[
\alpha =n+\frac{1}{2} \ \ \ with\ \ n=\{%
\begin{array}{c}
0,1,2,..................Re(E)\succ 0 \\
-1,-2,-3,..........Im(E)\prec 0%
\end{array}%
\]

$\Theta_{O}^{(n)}$ denotes the n$^{th}$ derivatives of $\Theta$
evaluated at $r_{0}$, the value of $r$ at which $V$ attains maximum.
Here a complexity arises from the fact that the potential $V$ is a
function of frequency $\omega$. This makes difficult to calculate
the value of $r_{0}$. We make use of the procedure suggested by
Konoplya\cite{ref6} to find $r_{0}$ by fixing all the parameters
other than $\omega$ for which the $V$ depends and then find the
value of $r$ at which $V$ attains maximum as a numerical function of
$\omega$. Substitute for $r_{0}$ and then find the value of $\omega$
which satisfies the condition (\ref{eqn9}) by trial and error way.

\section{Results and Discussion}
\begin{figure}[h]
\centering
\includegraphics[width=0.6\columnwidth]{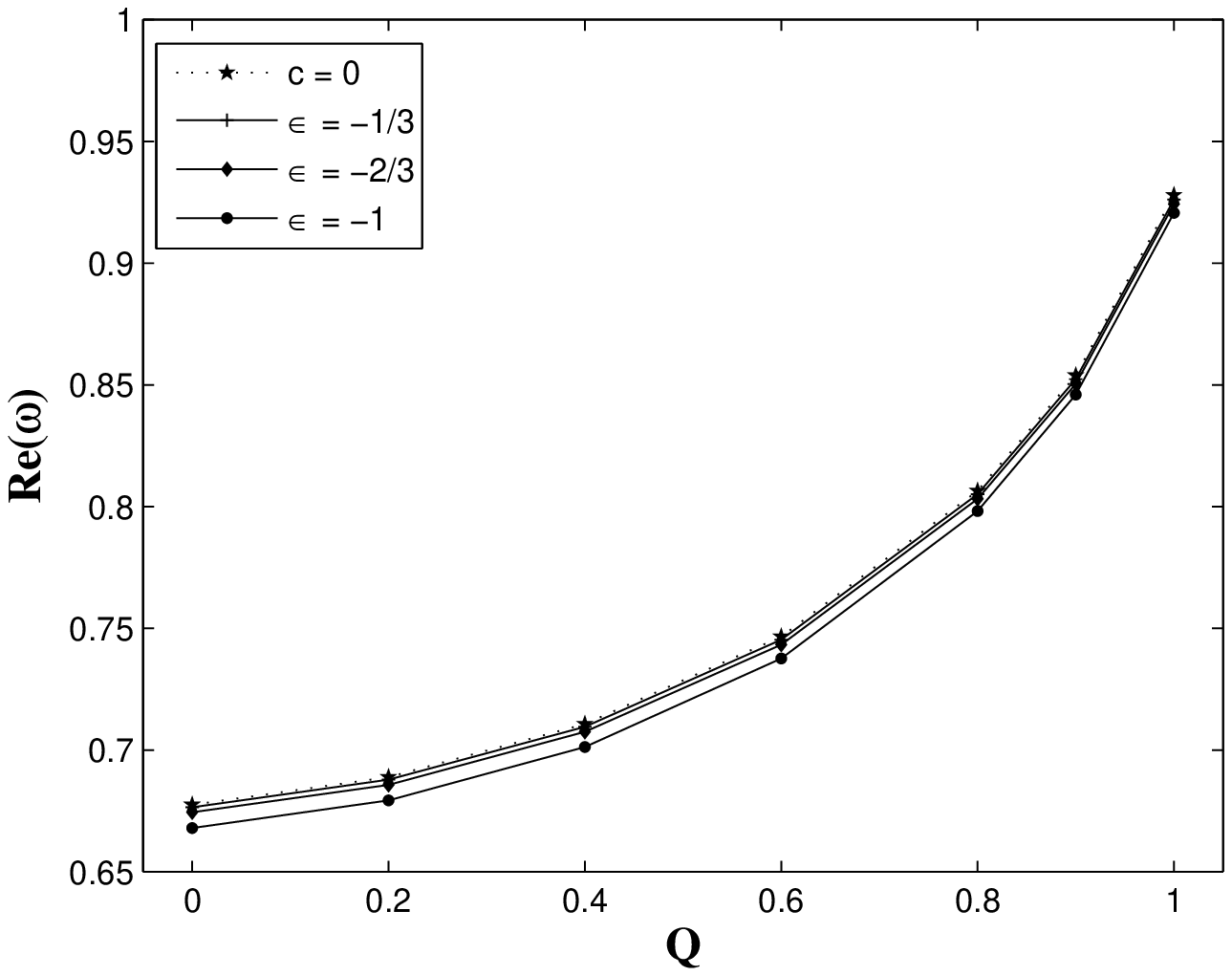}%
\hspace{0.1in}%
\includegraphics[width=0.6\columnwidth]{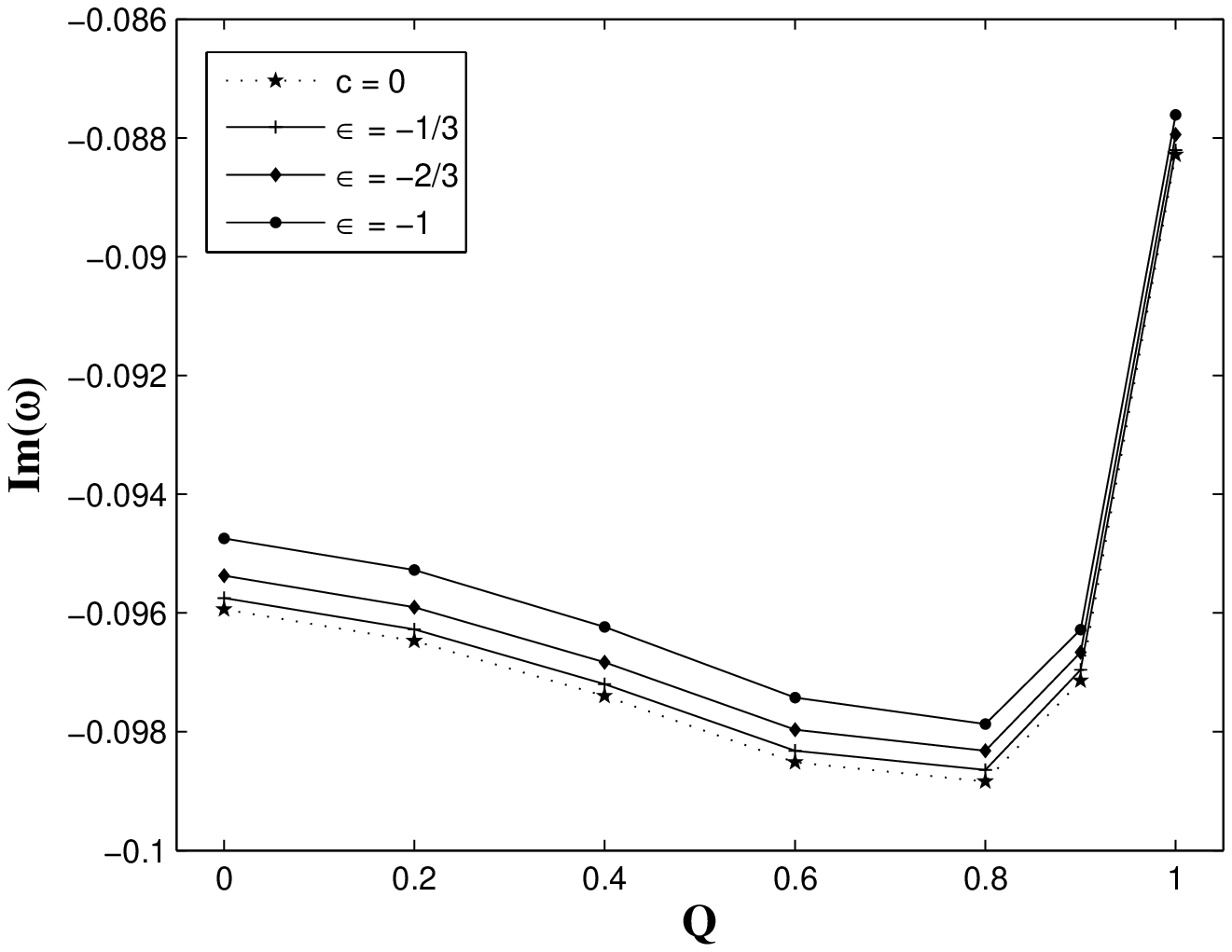}
\caption{$Re\omega$ and $Im\omega$ as a function of $Q$ for
$l=3,n=0,e=0.1,m=0.1$ and for different values of $\epsilon$ with
$c=0.001$. The dotted line represents the no quintessence case(c=0).
} %
\label{fig1}
\end{figure}

We take all the values of parameters in black hole mass(M) units. It
is a general experience that the imaginary part of quasinormal
frequencies decreases with increase in mode number, $n$, means that
the quasinormal frequencies with lower mode number will decay slowly
and are relevant to the description fields around black hole. So we
consider frequencies for lower mode number for our study.  The
dependence of real and imaginary parts of $\omega$ on the charge of
the black hole, $Q$ is plotted in Figure\ref{fig1} for fixed
$l=3,n=0,e=0.1,m=0.1$ and for different values of $\epsilon $ with
$c=0.001$. The case with absence of quintessence is also plotted
(dotted line).

The quasinormal frequencies for scalar field in charged black hole
is influenced by quintessence. The magnitudes of real and imaginary
parts of $\omega$ is lower in the presence of quintessential field.
This implies due to the presence of quintessence, the quasinormal
frequencies for scalar field in RN black hole damps more slowly. As
in the absence of quintessence\cite{ref6,ref8}, $Re(\omega)$
increases monotonically with the increase in $Q$ while the magnitude
of $Im(\omega)$ first decreases, falling to a minimum around $Q=0.8$
and thereafter increases sharply. For larger values of $Q$, the
modification with quintessence decreases.

\begin{figure}[h]
\centering
\includegraphics[width=0.45\columnwidth]{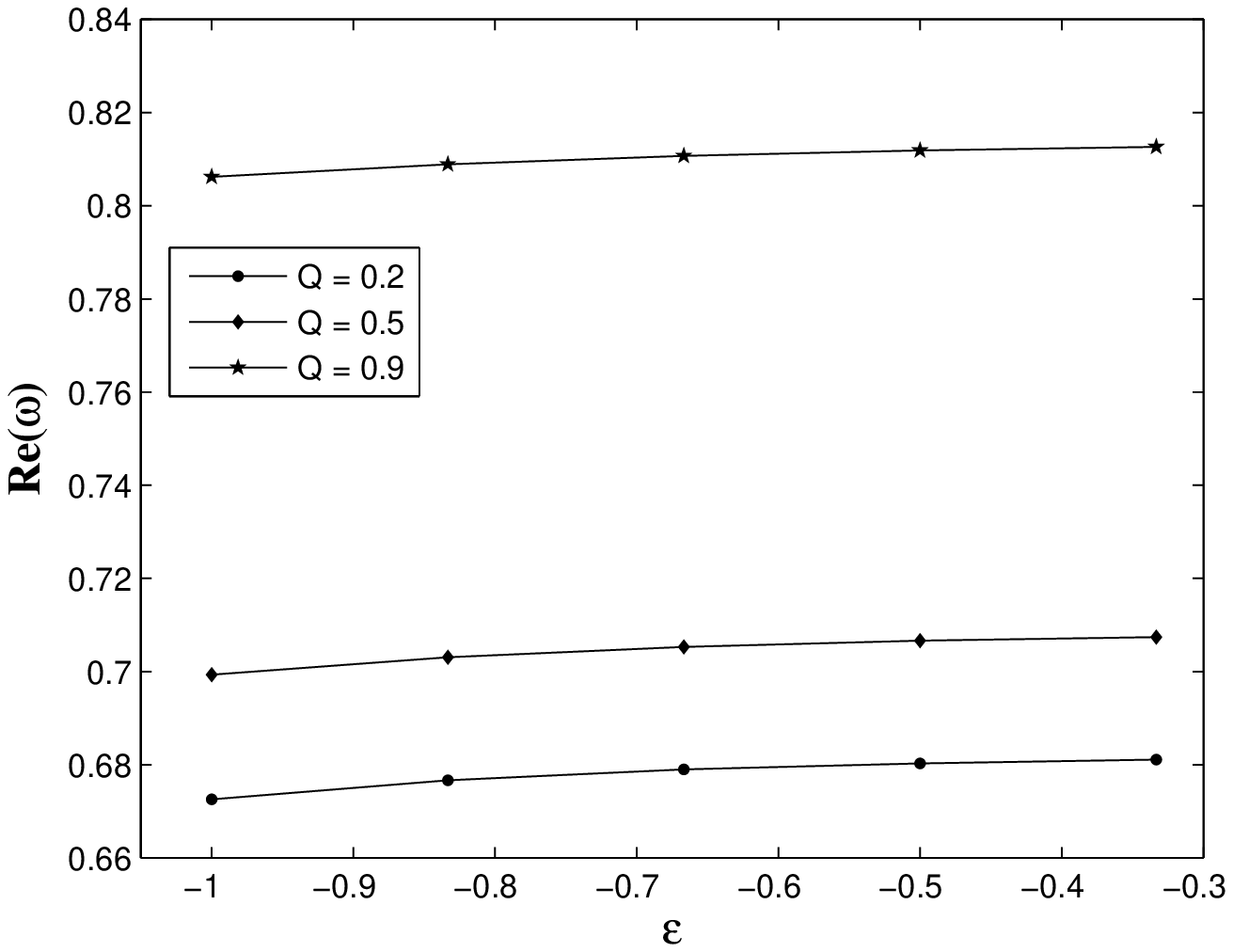}%
\hspace{0.1in}%
\includegraphics[width=0.45\columnwidth]{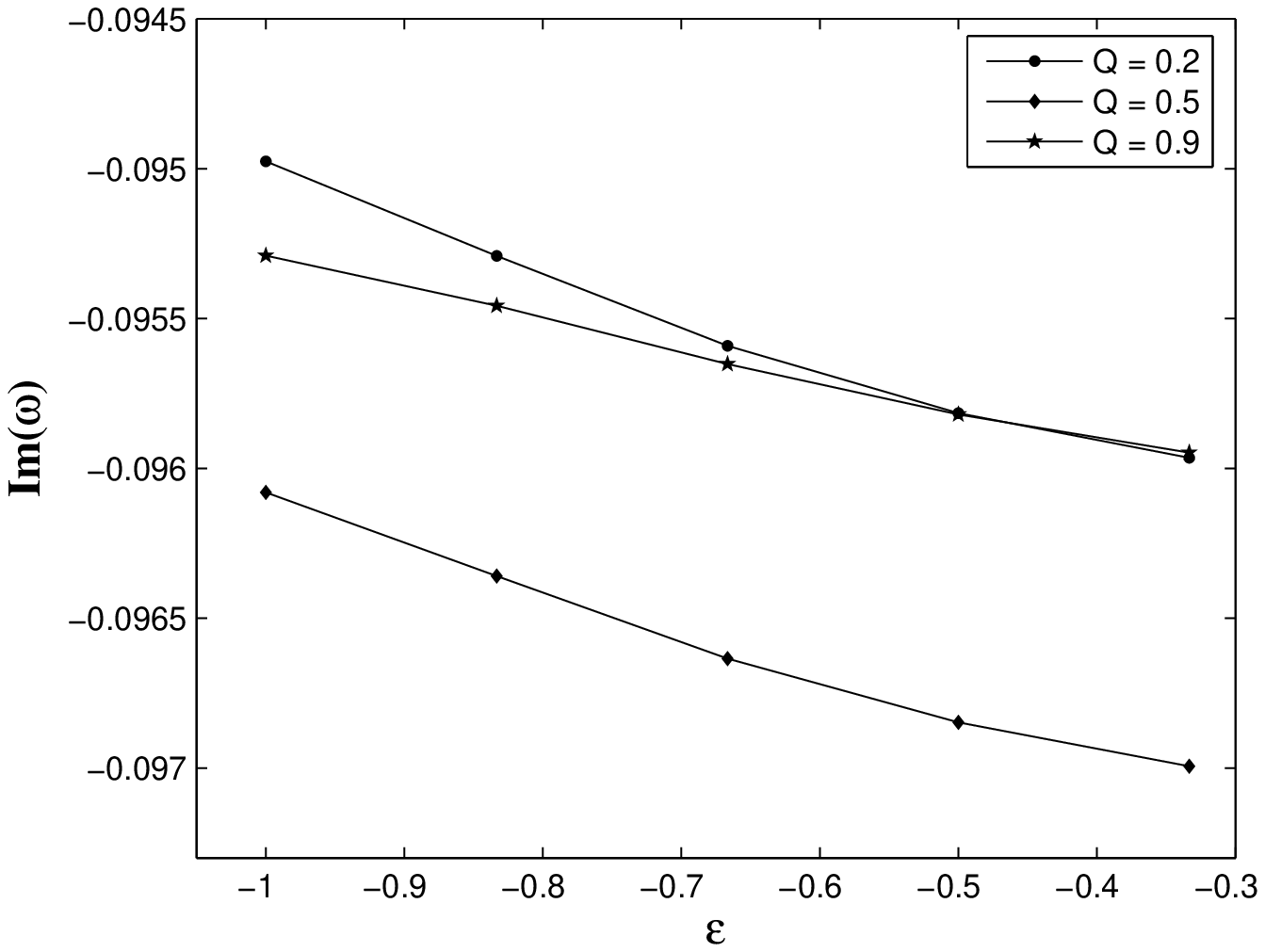}
\caption{Variation of $Re\omega$ and $Im\omega$ with $\epsilon$ for
$l=3,n=0,e=0,m=0.1$} %
\label{fig2}
\end{figure}

Figure\ref{fig2} shows the variation of $Re\omega$ and $Im\omega$
with quintessential state parameter $\epsilon$ with $c=0.001$ for
fixed $l=3,n=0,e=0,m=0.1$. As the value of $\epsilon$ increases
$Re\omega$ increases slowly but the magnitude of $Im\omega$
increases more fast, means damping is less for lower values of
$\epsilon$.

\begin{figure}[h]
\centering
\includegraphics[width=0.45\columnwidth]{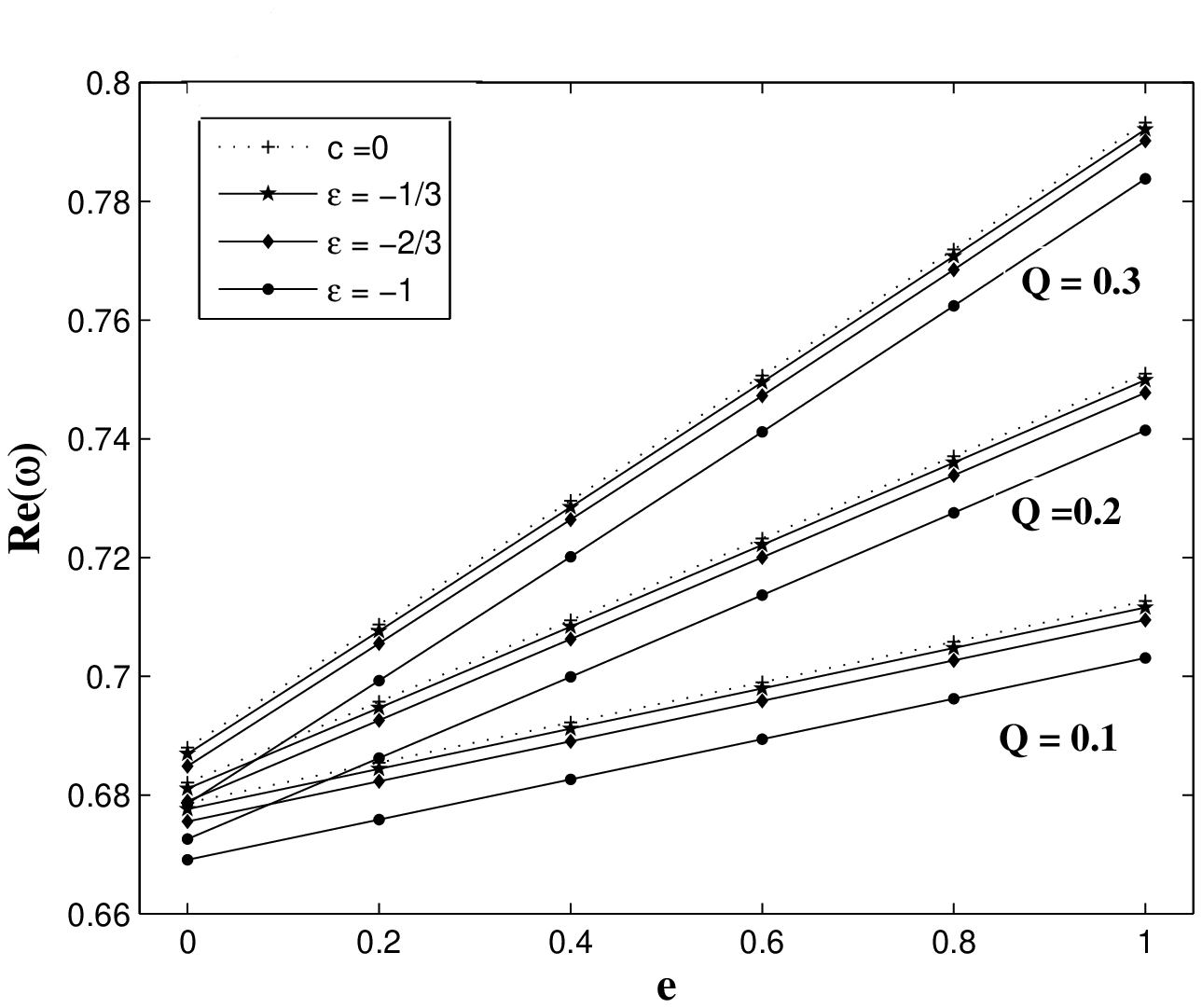}%
\hspace{0.1in}%
\includegraphics[width=0.45\columnwidth]{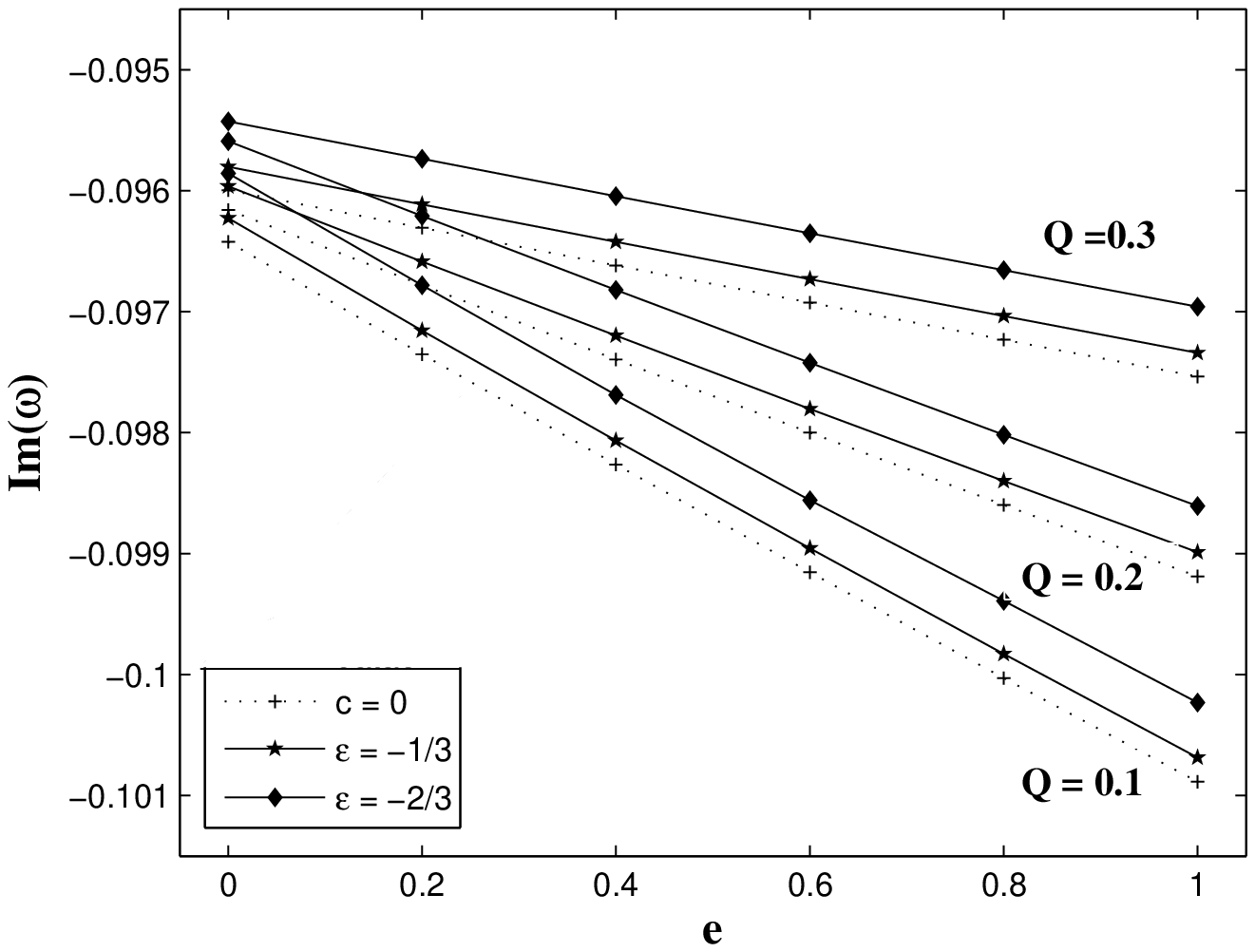}
\caption{Variation of $Re\omega$ and $Im\omega$ with $e$ for
$l=3,n=0,m=0.1,Q=0.1,0.2,0.3$ and different values of $\epsilon$. Dotted line is for $c=0$.} %
\label{fig3}
\end{figure}

In Figure\ref{fig3} $Re\omega$ and $Im\omega$ are plotted as
functions of $e$ with $l=3,n=0,m=0.1$ for $Q=0.1,0.2,0.3$ and
different values of $\epsilon$. Dotted line represents absence of
quintessence(c=0). The variation is almost linear. In the presence
of quintessence, the magnitude of  $Re\omega$ and $Im\omega$
increases with $e$ imitating the no quintessence case but with lower
values of $Re\omega$ and $|Im\omega|$. The modification with $e$
becomes larger for higher values of $Q$ .

Finally, we study the role of mass of scalar field on quasinormal
frequencies. Quasinormal modes occur only when the peak value of the
potential $V(r=r_{max})$ is larger than $m^2$ and $\omega^2$ of the
field is smaller than this peak value\cite{ref7}. This means that
there exists a maximum value for mass, $m_{max}$ beyond which there
will be no quasinormal modes. $m_{max}$ can be estimated from the
condition for the existence of quasinormal modes,

\begin{equation}
\ V(r_{\max ,}\omega =m_{\max })=(m_{\max })^{2} \label{eqn12}
\end{equation}

The values of $m_{max}$ obtained for different values of $\epsilon$
is tabulated in Table\ref{table1}. In the presence of quintessence
$m_{max}$ decreases  because quintessence lowers the height of the
potential barrier as shown in Figure\ref{fig5} and when
$\epsilon=-1$, it has the lowest value.

\begin{table}[h]
\centering
\begin{tabular}{|l|l|l|l|l|l|}
\hline $c$ & $\epsilon$ & $m_{max}$ & $c$ & $\epsilon$ & $m_{max}$
\\ \hline
0 & -- & \-0.88516 & 0.001 & -2/3 & 0.87815 \\
0.001 & -1/3 & 0.88366 & 0.001 & -1 & \-0.85864\\\hline
\end{tabular}%
\caption{The limit of mass of scalar field, $m_{max}$ for the
existence of quasinormal frequencies with $l=3,e=0,Q=0.1$}
\label{table1}
\end{table}

\begin{figure}[h]
\centering
\includegraphics[width=0.45\columnwidth]{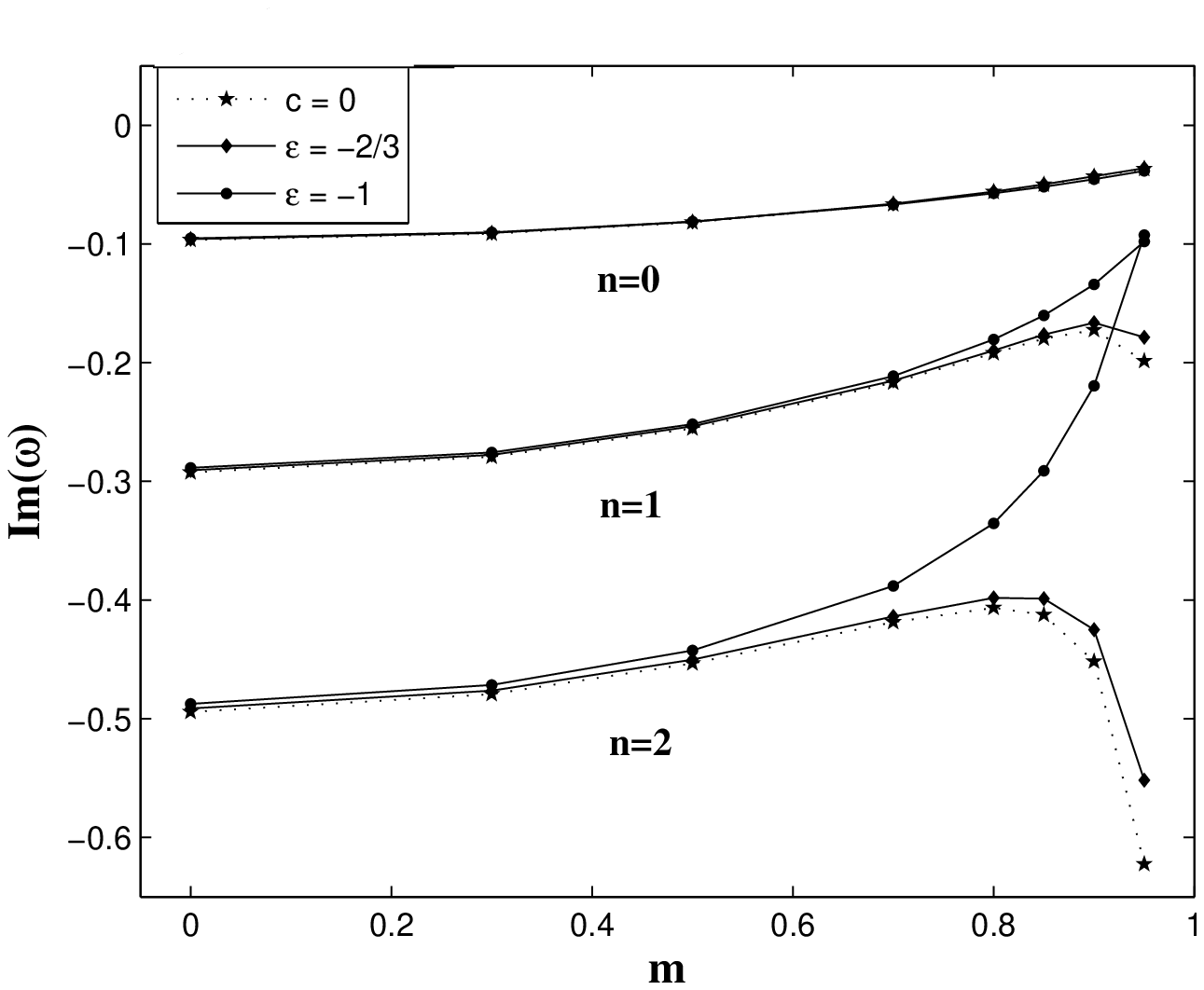}%
\hspace{0.1in}%
\includegraphics[width=0.45\columnwidth]{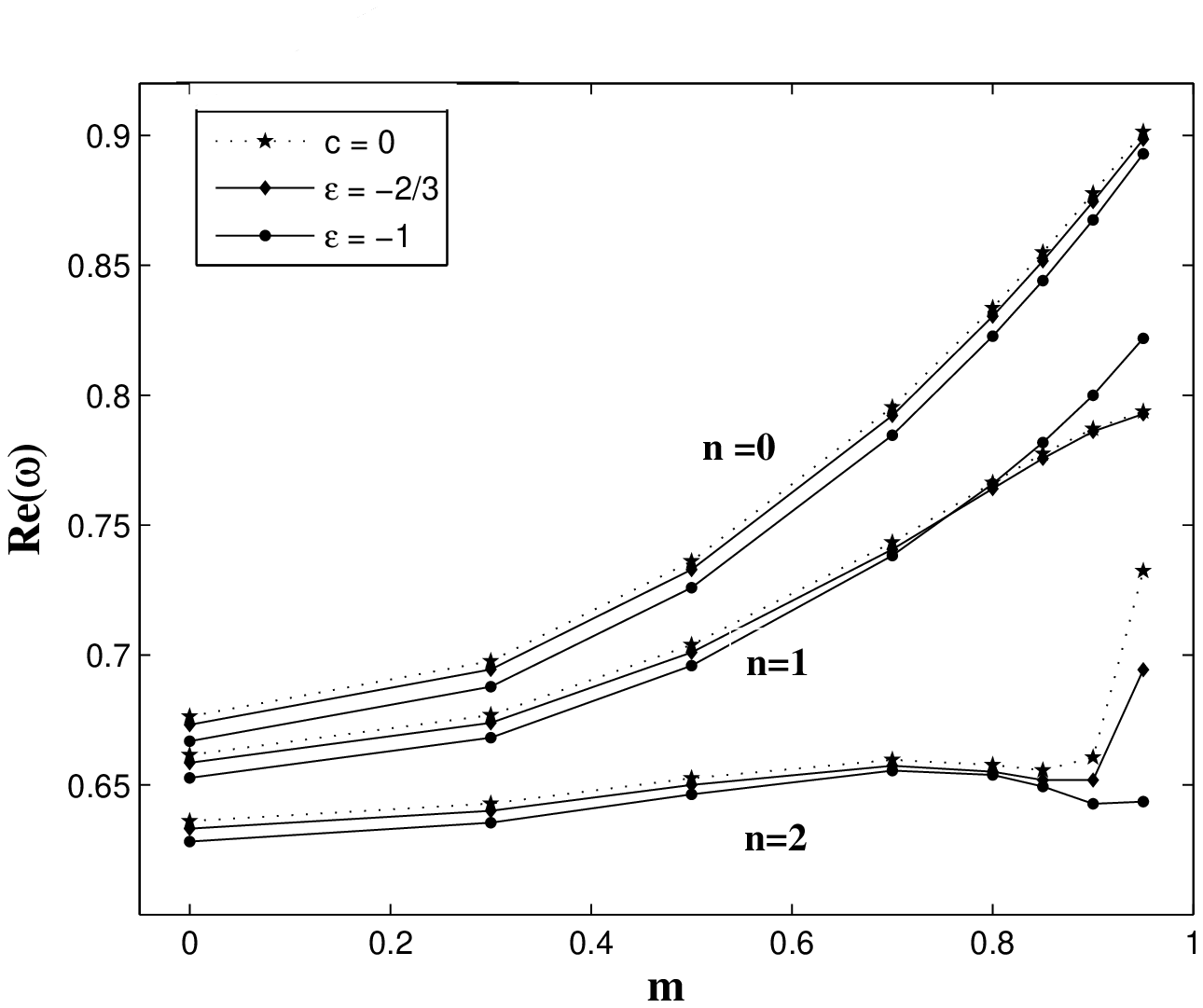}
\caption{Variation of $Re\omega$ and $Im\omega$ with $e$ for
$l=3,n=0,m=0.1,Q=0.1,0.2,0.3$ and different values of $\epsilon$. Dotted line is for $c=0$.} %
\label{fig4}
\end{figure}

\begin{figure}[h]
\centering
\includegraphics[width=0.6\columnwidth]{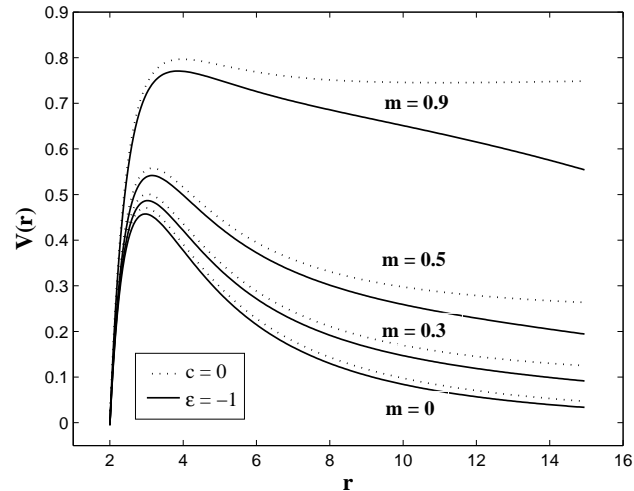}%
\caption{Effective potential
$l=3,n=0,m=0.1,Q=0.1,0.2,0.3$ and different values of $\epsilon$. Dotted line is for $c=0$.} %
\label{fig5}
\end{figure}

The values of  $Re\omega$ and  $Im\omega$ evaluated using WKB method
with the variation of mass is plotted in Figure\ref{fig4}.

$Re\omega$ increases with increase in mass while $|Im\omega|$
decreases with mass, which indicates that QNMs of fields with lower
mass damp slowly. But they behave abnormally near $m_{max}$. This is
due to the fact that, for larger mass of the field the potential
looses its barrier shape by broadening the potential peak as shown
in Figure\ref{fig5} and WKB method gives inaccurate results. Lower
modes show less abnormality showing that WKB method is more accurate
for fundamental modes.

Another thing we noticed is that this abnormal behavior is lower in
the presence of quintessence and even when $\epsilon=-1$ we can get
satisfactory curve because the peak of effective potential broaden
much less in the presence of quintessence comparing with the normal
case(c=0) as understood from Figure\ref{fig5} and quintessence helps
to retain barrier shape and give more accurate results at larger
mass range.
\section{Acknowledgment}
VCK is thankful to U.G.C, New Delhi for financial support through a
Major Research Project and wishes to acknowledge Associateship of
IUCAA, Pune, India.

\end{document}